\documentstyle[aps,preprint,amssymb,epsf]{revtex}
\begin{document}

\title{Cauchy-characteristic extraction in numerical relativity}

\author{Nigel T. Bishop${}^{1}$,
        Roberto G\'omez${}^{2}$,
	Luis Lehner${}^{2}$ and
        Jeffrey Winicour${}^{2}$
	}
\address{
${}^{1}$Department of Mathematics, Applied Mathematics and Astronomy,\\
University of South Africa, P.O. Box 392, Pretoria 0001, South Africa \\
${}^{2}$Department of Physics and Astronomy,\\
University of Pittsburgh, Pittsburgh, PA 15260}

\maketitle

\begin{abstract}

We treat the calculation of gravitational radiation using the mixed
timelike-null initial value formulation of general relativity. The
determination of an exterior radiative solution is based on boundary
values on a timelike worldtube $\Gamma$ and on characteristic data on
an outgoing null cone emanating from an initial cross-section of
$\Gamma$. We present the details of a 3-dimensional computational
algorithm which evolves this initial data on a numerical grid, which is
compactified to include future null infinity as finite grid points. A
code implementing this algorithm is calibrated in the quasispherical
regime. We consider the application of this procedure to the extraction
of waveforms at infinity from an interior Cauchy evolution, which
provides the boundary data on $\Gamma$. This is a first step towards
Cauchy-characteristic matching in which the data flow at the boundary
$\Gamma$ is two-way, with the Cauchy and characteristic computations
providing exact boundary values for each other. We describe strategies
for implementing matching and show that for small target error it is
much more computationally efficient than alternative methods.

\end{abstract}

\section{Introduction}

We report here on an important step towards the ultimate goal of
constructing numerical relativity codes that calculate accurately in 3D
the gravitational radiation at future null infinity. By ``accurately''
we mean (at least) second-order convergent to the true analytic
solution of a well posed initial value problem. Thus our goal is to
provide an accurate and unambiguous computational map from initial data
to gravitational waveforms at infinity. Of course, uncertainties will
always exist in the appropriate initial data for any realistic
astrophysical system (e.g. in a binary neutron star system, the data
for the metric components would not be uniquely determined by
observations). But such a computational map enables focusing on the
underlying physics in a rigorous way.

Most relativity codes are second-order convergent, but because of
boundary problems the convergence may not be  to the true analytic
solution of the intended physical problem. In order to explain this
point, and to give the idea behind our method, we first briefly review
some aspects of numerical relativity.  The predominant work in
numerical relativity is for the Cauchy ``3 + 1'' problem, in which
spacetime is foliated into a sequence of spacelike hypersurfaces. These
hypersurfaces are necessarily of finite size so, in the usual case
where space is infinite, an outer boundary with an artificial boundary
condition must be introduced. This is the first source of error because
of artificial effects such as the reflection of outgoing waves by the
boundary.  Next, the gravitational radiation is estimated from its form
inside the boundary by using perturbative methods, which ignore the
nonlinear aspects of general relativity in the region outside the
boundary. For these reasons the numerical estimate of gravitational
radiation is not, in general, convergent to the true analytic value at
future null infinity.  The radiation properties of the
Robinson-Trautman metric will be used to illustrate this effect.

An alternative approach in numerical relativity uses the characteristic
formalism, in which spacetime is foliated into a sequence of null cones
emanating from a central geodesic.  This approach has the advantage that
the Einstein equations can be compactified~\cite{null-infinity} so that
future null infinity is rigorously represented on a finite grid, and
there is no artificial outer boundary condition. However, it suffers
from the disadvantage that the coordinates are based on light rays,
which can be focussed by a strong field to form caustics which
complicate a numerical computation~\cite{Stewart}. Also, to date, the
characteristic initial value problem has only been implemented
numerically for special symmetries
{}~\cite{IWW,piran87,Bis90,Stew,Clarke,papa}.

Our ultimate goal is a 3D Cauchy-characteristic matching (CCM) code,
which uses the positive features of the two methods while avoiding the
problems.  More precisely, the interior of a timelike worldtube
$\Gamma$ is evolved by a Cauchy method, and the exterior to future null
infinity is evolved using a characteristic algorithm;  boundary
conditions at $\Gamma$ are replaced by a two-way flow of information
across $\Gamma$.  In relativity, under the assumption of axisymmetry
without rotation, there has been a feasibility study of
CCM~\cite{Bis,Bis2}; see also~\cite{Clarke}.  CCM has been successfully
implemented for nonlinear wave equations and demonstrated to be
second-order convergent to the true analytic solution (which is not
true in a pure Cauchy formulation with Sommerfeld outer boundary
condition)~\cite{Ccpaper}.

While CCM has aesthetic advantages, it is important to ask whether it
is an efficient approach. The question can be posed as follows.  For a
given target error $\varepsilon$, what is the amount of computation
required for CCM compared to that required for a pure Cauchy
calculation?  It will be shown that the ratio tends to $0$ as
$\varepsilon \rightarrow 0$, so that in the limit of high accuracy the
effort is definitely worthwhile~\cite{effic}.

Our first step towards CCM is Cauchy-characteristic extraction (CCE)
and we will present a partial implementation of CCE in this paper. The
idea of CCE is to run a pure Cauchy evolution with an approximate outer
boundary condition. A worldtube $\Gamma$ is defined in the interior of
the Cauchy domain, and the appropriate characteristic data is
calculated on $\Gamma$; then characteristic algorithms are used to
propagate this gravitational field to future null infinity~\cite{tam}.
CCE is simpler than CCM to implement numerically, because in CCE the
data flow is one-way (Cauchy to characteristic) whereas in CCM the data
flows in both directions. Note that the advantage of computational
efficiency applies only to CCM and not to CCE. However, we will show
that the advantage of second-order convergence to the true analytic
solution does apply, under certain circumstances, to CCE.

The work in this paper is part of the Binary Black Hole Grand
Challenge, which is concerned with the gravitational
radiation resulting from the in-spiral and coalescence of two arbitrary
black holes. However, the methods described here are not limited to
black hole coalescence and could be applied to gravitational radiation
from any isolated system, either with or without matter.

In Sec.~\ref{sec:2}, we present a formalism for 3D characteristic
numerical relativity in which the coordinates are based on null cones
that emanate from a timelike worldtube $\Gamma$ (Recall that existing
codes are in 2D with null cones emanating from a timelike geodesic)
{}~\cite{papa}.
The characteristic Einstein equations are written as a sum of two
parts:  quasispherical (in a sense defined below) plus nonlinear. The
discretization and compactification of the Einstein equations, with the
nonlinear part ignored, is discussed in Sec.~\ref{sec:3}.
A computer code has been written and in Sec.~\ref{sec:4} this code
is tested on linearized solutions of the Einstein equations, and
extraction is tested on the nonlinear Robinson-Trautman solutions. The
Robinson-Trautman solutions are also used to investigate the error of
perturbative methods in estimating the gravitational radiation at null
infinity. Sec.~\ref{sec:5} uses the formalism developed in
Sec.~\ref{sec:2} to estimate the errors associated with the finite
boundary in a pure Cauchy computation. This leads to the result
concerning computational efficiency of CCM stated above.  In the
Conclusion we discuss the further steps needed for a full
implementation of CCE, and also of CCM, and investigate under what
circumstances CCE can provide second-order convergence to the true
analytic solution at future null infinity.  We finish with Appendices
on the null cone version of gauge freedom and linear solutions of the
Einstein equations, and on a stability analysis of our algorithm.

\section{Characteristic evolution in 3D} \label{sec:2}

This is the first step towards a 3D characteristic evolution algorithm
for the fully nonlinear vacuum Einstein equations.  Here we treat the
quasispherical case, where effects which are nonlinear in the asymmetry
can be ignored. Thus the Schwarzschild metric is treated exactly in
this formalism. However, rather than developing an algorithm for the
linearized equations on a given Schwarzschild background, we will
approach this problem in a mathematically different way.

We adopt a metric based approach in which each component of Einstein's
equation has (i) some quasispherical terms which survive in the case of
spherical symmetry and (ii) other terms which are
quadratic in the asymmetry, i.e terms of $O(\lambda^2)$ where $\lambda$
measures deviation from spherical symmetry. We will treat the
quasispherical terms to full nonlinear accuracy while discarding the
quadratically asymmetric terms. For example, if $\phi$ were a scalar
function we would make the approximation
\begin{equation}
e^{\phi} \partial_{\theta}^2 e^{\phi}
+ \partial_{\theta} \phi \partial_{\theta} \phi \approx
e^{\phi} \partial_{\theta}^2 e^{\phi}
\end{equation}

Although this breakup is not unique, once made it serves two useful
purposes. First, the resulting  field equations are physically
equivalent to the linearized Einstein equations in the quasispherical
regime. (In the exterior vacuum region, the spherical background must
of course be geometrically Schwarzschild but the quasispherical
formalism maintains arbitrary gauge freedom in matching to an interior
solution). Second, the resulting quasispherical evolution algorithm
supplies a building block which can be readily expanded into a fully
nonlinear algorithm by simply inserting the quadratically asymmetric
terms in the full Einstein equations.

\subsection{The null cone formalism}

We use coordinates based upon a family of outgoing null hypersurfaces.
We let $u$ label these hypersurfaces, $x^A$ ($A=2,3$), be labels for
the null rays and $r$ be a surface area distance. In the resulting
$x^\alpha=(u,r,x^A)$ coordinates, the metric takes the Bondi-Sachs
form~\cite{bondi,sachs}
\begin{equation}
   ds^2=-\left(e^{2\beta}{V \over r} -r^2h_{AB}U^AU^B\right)du^2
        -2e^{2\beta}dudr -2r^2 h_{AB}U^Bdudx^A +r^2h_{AB}dx^Adx^B,
   \label{eq:bmet}
\end{equation}
where $h^{AB}h_{BC}=\delta^A_C$ and $det(h_{AB})=det(q_{AB})=q$, with
$q_{AB}$ a unit sphere metric. Later, for purposes of including null
infinity as a finite grid point, we introduce a compactified
radial coordinate.

A Schwarzschild geometry is given by the choice $\beta=\beta(u)$,
$V=e^{2\beta}(r-2m)$, $U_A=0$ and $h_{AB}=q_{AB}$. To describe a linear
perturbation, we would set $h_{AB}=q_{AB}+\lambda \gamma_{AB}$ and
would retain only terms in $\gamma_{AB}$ which were of leading order in
the linearization parameter $\lambda$. Here we take a different
approach. We express
\begin{equation}
    q_{AB} =\frac{1}{2}\left(q_A \bar q_B+\bar q_Aq_B\right),
\end{equation}
in terms of a complex dyad $q_A$ (satisfying $q^Aq_A=0$, $q^A\bar q_A
=2$, $q^A=q^{AB}q_B$, with $q^{AB}q_{BC}=\delta^A_C$). Then the dyad
component $J=h_{AB}q^Aq^B /2$ is related to the linearized metric by
$J=\lambda\gamma_{AB}q^Aq^B /2$. In linearized theory, $J$ would be a
first order quantity. The 2-metric is uniquely determined by $J$, since
the determinant condition implies that the remaining dyad component
$K=h_{ab}q^a \bar q^b /2$ satisfies $1=K^2-J\bar J$. Refer to
{}~\cite{eth} for further details, especially how to discretize the
covariant derivatives and curvature scalar of a topologically spherical
manifold using the $\eth$ calculus.

Because the 2-metric also specifies the null data for the
characteristic initial value problem, this role can be transferred to
$J$. Terms in Einstein equations that depend upon $J$ to higher than
linear order are quadratically asymmetric. We do not explicitly
introduce $\lambda$ as a linearization parameter but introduce it where
convenient to indicate orders of approximation.

\subsection{Quasispherical approximation} \label{sec:2.2}

The Einstein equations $G_{\mu\nu}=0$ decompose into hypersurface
equations, evolution equations and conservation laws. In writing the
field equations, we follow the formalism given in~\cite{newt,nullinf}.
We find:
\begin{eqnarray}
\beta_{,r} &=& \frac{1}{16}rh^{AC}h^{BD}h_{AB,r}h_{CD,r},
\label{eq:beta}
\\
(r^4e^{-2\beta}h_{AB}U^B_{,r})_{,r}  &=&
2r^4 \left(r^{-2}\beta_{,A}\right)_{,r}
-r^2 h^{BC}D_{C}h_{AB,r}
\label{eq:u}
\\
2e^{-2\beta}V_{,r} &=& {\cal R} - 2 D^{A} D_{A} \beta
-2 D^{A}\beta D_{A}\beta + r^{-2} e^{-2\beta} D_{A}(r^4U^A)_{,r}
-\frac{1}{2}r^4e^{-4\beta}h_{AB}U^A_{,r}U^B_{,r},
\label{eq:v}
\end{eqnarray}
where $D_A$ is the covariant derivative and ${\cal R}$ the curvature
scalar of the 2-metric $h_{AB}$.

The quasispherical version of (\ref{eq:beta}) follows immediately from
rewriting it as $\beta_{,r}=N_{\beta}$, where
$N_{\beta}=rh^{AC}h^{BD}h_{AB,r}h_{CD,r}/16$ is quadratically
asymmetric. This defines the quasispherical equation
\begin{equation}
\beta_{,r}= 0.
\label{eq:wbeta}
\end{equation}
Thus in this approximation, $\beta=H(u,x^A)+O(\lambda^2)$. For
a family of outgoing null cones which emanate from a
nonsingular geodesic worldline, we could choose coordinate conditions so
that $H=0$. Similarly, in Minkowski space, we could set $H=0$ for null
hypersurfaces which emanate from a non-accelerating spherical worldtube of
constant radius. In a Schwarzschild spacetime, due to red shift
effects, $H$ need not vanish even on a spherically symmetric
worldtube.  Thus $H$ represents some physical information as well as
asymmetric gauge freedom in the choice of coordinates and choice of
world tube.

We wish to apply the same procedure to equations (\ref{eq:u}) and
(\ref{eq:v}). In doing so, it is useful to introduce the $O(\lambda)$
tensor field
\begin{equation}
C^C_{AB}={1\over 2} h^{CD}(\nabla_A h_{DB}+\nabla_B h_{AD}-\nabla_D h_{AB})
\label{eq:christoffel}
\end{equation}
which represents the difference between the connection $D_A$ and the
unit sphere connection $\nabla_A$, e.g.
$(D_A-\nabla_A)v_B=-C^C_{AB}v_C$. In solving for $U^A$, we use the
intermediate variable
\begin{equation}
Q_A = r^2 e^{-2\,\beta} h_{AB} U^B_{,r}.
\end{equation}

Then (\ref{eq:u}) reduces to the first order radial equations
\begin{eqnarray}
\left(r^2 Q_A\right)_{,r} &=& 2r^4 \left(r^{-2} \beta_{,A} \right)_{,r}
-r^2 h^{BC} D_{C} h_{AB,r} ,
\label{eq:Qa}
\\
U^A_{,r} &=& r^{-2} e^{2\beta} h^{AB}Q_B.
\label{eq:ua}
\end{eqnarray}
We deal with these equations in terms of the spin-weighted fields
$U=U^Aq_A$ and $Q=Q_Aq^A$.  To obtain quasispherical versions of these
equations, we rewrite
(\ref{eq:Qa}) and (\ref{eq:ua}) as
\begin{eqnarray}
(r^2 Q)_{,r} &=& -r^2q^{A}q^{BC}\nabla_{C}h_{AB,r}
+2 r^4 q^A \left(r^{-2} \beta_{,A}\right)_{,r} + N_Q,
\label{eq:wQa}
\\
U_{,r} &=& r^{-2}e^{2\beta}Q + N_U ,
\label{eq:wua}
\end{eqnarray}
where
\begin{eqnarray}
N_Q &=& q^{A}\left[r^2h^{BC} \left(C^D_{CA}h_{DB,r} +C^D_{CB}h_{AD,r}\right)
-r^2 \left(h^{BC}-q^{BC}\right) \nabla_{C}h_{AB,r}\right] ,
\\
N_U &=& r^{-2}e^{2\beta}q_{A} \left(h^{AB}-q^{AB}\right) Q_B.
\end{eqnarray}
The quasispherical versions obtained by setting $N_Q=0$ in
(\ref{eq:wQa}) and $N_U=0$ in (\ref{eq:wua}) then take the form
\begin{eqnarray}
(r^2 Q)_{,r} &=& -r^2 (\bar \eth J + \eth K)_{,r}
+2r^4\eth \left(r^{-2}\beta\right)_{,r} ,
\label{eq:wq}
\\
U_{,r} &=& r^{-2}e^{2\beta}Q ,
\label{eq:wu}
\end{eqnarray}
in terms of the spin-weighted differential operator $\eth$. Since
$Q_{,r}$ and $U_{,r}$ are asymmetric of $O(\lambda)$, we use the gauge
freedom to ensure that $Q$ and $U$ are $O(\lambda)$.

Since $V=r$ in Minkowski space, we set $V=r+W$ in terms of a
quasispherical variable $W$. Then (\ref{eq:v}) becomes
\begin{equation}
W_{,r}=\frac{1}{2} e^{2\beta}{\cal R} -1
- e^{\beta} \eth \bar \eth e^{\beta}
+ \frac{1}{4} r^{-2} \left(r^4
\left(\eth \bar U +\bar \eth U \right)
\right)_{,r} + N_W ,
\label{eq:ww}
\end{equation}
where
\begin{equation}
N_W= - e^{\beta} \nabla_A \left[ \left(h^{AB}-q^{AB}\right)
\nabla_B e^{\beta}\right]
- \frac{1}{4}r^4 e^{-2\beta} h_{AB} U^A_{,r} U^B_{,r}.
\label{eq:nw}
\end{equation}
We set $N_W=0$ in (\ref{eq:ww}) to obtain the quasispherical field
equation for $W_{,r}$.

Next, by the same procedure, the evolution equations take the form
\begin{equation}
2 \left(rJ\right)_{,ur} - \left(r^{-1}V\left(rJ\right)_{,r}\right)_{,r}
= -r^{-1} \left(r^2\eth U\right)_{,r} + 2 r^{-1} e^{\beta} \eth^2 e^{\beta}
- \left(r^{-1} W \right)_{,r} J + N_J ,
\label{eq:wev}
\end{equation}
where
\begin{eqnarray}
 N_J= {{q^A q^B} \over r} [
    &-&2 e^{\beta} C^C_{AB} \nabla_C e^{\beta}
    - h_{AC} C^C_{DB} \left(r^2U^D\right)_{,r}
    - \left(h_{AC}-q_{AC}\right) \nabla_B \left(r^2U^C\right)_{,r}
    + \frac{1}{2} r^4 e^{-2\beta} h_{AC} h_{BD} U^C_{,r} U^D_{,r}
\nonumber \\
    &-& \frac{1}{2} r^2 h_{AB,r} D_CU^C -r^2 U^C D_C h_{AB,r}
    +r^2 h^{CD} h_{AD,r} \left(D_CU_B-D_BU_C\right) ] .
   \label{eq:nj}
\end{eqnarray}
The quasispherical evolution equation follows from (\ref{eq:wev}) by
setting $N_J=0$.

The remaining independent equations are the conservation conditions.
For a worldtube given by $r=constant$, these are given in terms of the
Einstein tensor by
\begin{equation}
\xi^{\mu}G_{\mu}^{\nu}\nabla_{\nu}r=0,
\end{equation}
where $\xi^{\mu}$ is any vector field tangent to the worldtube.  This
expresses conservation of $\xi$-momentum flowing across the
worldtube~\cite{tam}. These equations simplify when the Bondi
coordinates are adapted to the worldtube so that the angular
coordinates $x^A$ are constant along the $\partial_u$ streamlines. Then
$U=0$ on the worldtube and an independent set of conservation equations
is given (in the quasispherical approximation) in terms of the Ricci
tensor by
\begin{eqnarray}
   R_u^r =& r^{-2} W_{,u} - 2 r^{-1} \beta_{,u}
   - \frac{1}{2} r^{-3} \eth \bar \eth W
   + \frac{1}{4} \left(\eth \bar U + \bar \eth U\right)_{,r} &= 0 ,
\\
   2 q^A R_A^r  =& \eth \left( \left(r^{-1} W \right)_{,r}
   - 4 r^{-1}\beta - 2 \beta_{,u}\right)
   + \bar \eth \left(J_{,u} - J_{,r}\right)
   - r^2 U_{,ru} &= 0 .
\end{eqnarray}
In the context of an extraction problem it is assumed that the interior
solution satisfies the Einstein equations, and therefore that the
conservation conditions are automatically satisfied on the extraction
worldtube.

The above equations define a quasispherical truncation of the vacuum
Einstein equations. Because these quasispherical equations retain some
terms which are nonlinear in the asymmetry, their solutions are not
necessarily linearized solutions in a Schwarzschild background.
However, in the perturbative limit off Schwarzschild, the linearized
solutions to these truncated equations agree with the linearized
solutions to the full Einstein equations.

\section{Discretization of the Equations} \label{sec:3}

In this section we describe a numerical implementation, based on
second-order accurate finite differences, of the equations presented in
Sec.~\ref{sec:2}. We introduce a compactified radial coordinate
$x=r/(R+r)$, (with $R$ being the extraction radius), labeling null rays
by the real and imaginary parts of a stereographic coordinate
$\xi=q+ip$ on the sphere, i.e. $x^A=(q,p)$.  The radial coordinate is
discretized as $x_{i}=x_{0}+(i-1)\Delta x$ for $i=1\dots N_x$ and
$\Delta x = (1-x_{0}) / (N_{x}-1)$. Here $x_{0}=1/2$ defines a world
tube of constant surface area coordinate.  The point $x_{N_{x}}=1$ lies
at null infinity.  The stereographic grid points are given by
$q_{j}=j\Delta$ and $p_{k}=k\Delta$ for $j, k = -N_{\xi} \ldots
N_{\xi}$ and $\Delta = 1/N_{\xi}$.

The fields $J$, $\beta$, $Q$ and $W$ are represented by their values on
this rectangular grid, e.g.  $J^{n}_{ijk}=J(u_{n},x_i,q_{j},p_{k})$.
However, for stability (see Appendix \ref{app:stab}), the field $U$ is
represented by values at the points $x_{i+\frac{1}{2}} = x_{i}+\Delta
x/2$ on a radially staggered grid (accordingly
$U^{n}_{ijk}=U(u_{n},x_{i+\frac{1}{2}},q_{j},p_{k})$ ). For the extraction
problem, it is assumed that the values of the fields and the radial
derivative of $U$ are known at the boundary.  In the following
discussion, it is useful to note that asymptotic flatness implies that
the fields $\beta(x)$, $U(x)$, ${\tilde W(x)}= r^{-2} W(x)$ and $J(x)$
are smooth at $x=1$, future null infinity ${\cal I}^+$.

\subsection*{Hypersurface equation for $Q$}

In terms of the compactified radial variable $x$, the quasispherical
field equation for $Q$ reduces to
\begin{equation}
2 Q + x(1-x)Q_{,x} = -x(1-x)(\bar \eth J + \eth K)_{,x} - 4\eth\beta.
\label{eq:weakq}
\end{equation}
We write all derivatives in centered, second order accurate form and
replace the value $Q_{i-1}$ by its average $(Q_{i} + Q_{i-2})/2$. The
resulting algorithm determines $Q_{i}$ in terms of values of $J$ and
$\beta$ at the points $x_{i}$, $x_{i-1}$ and $x_{i-2}$
\begin{equation}
Q_{i} + Q_{i-2} + x_{i-1} (1 - x_{i-1}) {{Q_{i} - Q_{i-2}} \over {2 \Delta x}}
= - x_{i-1} (1 - x_{i-1})
\left( \bar\eth {{ J_{i} - J_{i-2} } \over {2 \Delta x} }
      + \eth    {{ K_{i} - K_{i-2} } \over {2 \Delta x} }
\right)
 - 4 \eth \beta_{i-1} .
\label{eq:qdisc}
\end{equation}
(Here and in what follows, we make explicit only the discretization on
the radial direction $x$, and we suppress the angular indices $j,k$).
Since Eq.~(\ref{eq:qdisc}) is a 3-point formula, it can not be applied at
the second point, however, a suitable formula for $x_{2}$ is given by
\begin{equation}
  Q_{i} + Q_{R} + x_{C} (1 - x_{C}) { {Q_{i} - Q_{R}} \over {\delta x} } =
  - x_{C} (1 - x_{C})
\left( \bar \eth { {J_{i} - J_{R} } \over {\delta x} }
          + \eth { {K_{i} - K_{R} } \over {\delta x} } \right)
  - 2 \eth \left( \beta_{i} + \beta_{R} \right) ,
\label{eq:qdisc2}
\end{equation}
where the value of $Q_{R}$ is trivially obtained from the knowledge of
$U_{,r}$ at the boundary, and $x_{C}=(x_{i} + x_{R})/2$, $\delta
x=x_{i}-x_{R}$.  After a radial march, the local truncation error
compounds to an $O(\Delta^2)$ global error in $Q$ at ${\cal I}^+$.

\subsection*{Hypersurface equation for $U$}

In terms of the compactified radial variable $x$, the quasispherical
field equation for $U$ reduces to
\begin{equation}
U_{,x} = {{e^{2\beta}Q} \over {R x^2} }.
\label{eq:weaku}
\end{equation}
We again rewrite all derivatives in centered, second order form.
Because of the staggered placement of $U$, the resulting discretization
is
\begin{equation}
U_i = U_{i-1} + \frac{e^{2\beta_i}Q_i}{ R x_i^2} \Delta x.
\label{eq:udisc}
\end{equation}
The value of $U$ at the first point is evaluated from the expansion
\begin{equation}
U_i = U|_R + U,_{x}|_R \left(x_{i+\frac{1}{2}} - x_R\right) + O(\Delta^2)
\end{equation}
at the boundary.  This leads to an algorithm for determining $U$ at the
point $x_{i+\frac{1}{2}}$ in terms of values of $Q$ at the points
$x_{R}$ lying on the same angular ray. After completing a radial march,
local truncation error compounds to an $O(\Delta^2)$ global error in
$U$ at ${\cal I^+}$.

\subsection*{Hypersurface equation for $W$}

The quasispherical field equation  for $W$ (\ref{eq:ww}), reexpressed
in terms of $x$ and $\tilde W = W/r^2$, is
\begin{equation}
R x^2 {\tilde W},_{x} + 2 R  {x \over {1 - x}} \tilde W =
\frac{1}{2} e^{2\beta} {\cal R} - 1 - e^{\beta} \eth \bar \eth e^{\beta}
+ \frac{1}{4} R x^2 \left( \eth \bar U + \bar \eth U \right)_{,x}
+ R {x \over {1 - x}} \left( \eth \bar U + \bar \eth U \right) .
\label{eq:wwxi}
\end{equation}
Following the same procedure as in Eq.~(\ref{eq:qdisc}) we obtain
\begin{eqnarray}
\lefteqn{
R x_{i-\frac{1}{2}}^2 (1-x_{i-\frac{1}{2}})
{ { \tilde W_{i} - \tilde W_{i-1} } \over \Delta x }
+ R x_{i-\frac{1}{2}} \left(\tilde W_i + \tilde W_{i-1} \right) = }
\nonumber \\
& & \frac{1}{2} (1-x_{i-\frac{1}{2}}) \left(
  \frac{1}{2} e^{2\beta_i}     {\cal R}_i
+ \frac{1}{2} e^{2\beta_{i-1}} {\cal R}_{i-1}
 - 2
 - e^{\beta_i}     \eth\bar \eth e^{\beta_{i}}
 - e^{\beta_{i-1}} \eth\bar \eth e^{\beta_{i-1}}
 \right) \nonumber \\
& &  + \frac{1}{4} R x_{i-\frac{1}{2}}^2 (1-x_{i-\frac{1}{2}})
\left( \eth { { \bar U_{i} - \bar U_{i-2} } \over {2 \Delta x} }
      + \bar \eth{ { U_{i} - U_{i-2} } \over {2 \Delta x} }
\right)
+ R x_{i-\frac{1}{2}}
  \left( \eth \bar U_{i-1} + \bar \eth U_{i-1} \right) .
\label{eq:wdisc}
\end{eqnarray}
We obtain a startup version of the above with the substitutions
$x_{i-\frac{1}{2}} \rightarrow x_{C}$, $\Delta x \rightarrow \delta x$,
noting that at the boundary $U,_{x}$ is given.  The above algorithm
has a local error $O(\Delta ^3)$in each zone.  In carrying out the
radial march, this leads to $O(\Delta ^2)$ error at any given physical
point in the uncompactified manifold. However, numerical analysis
indicates an $O(\Delta ^2 \log \Delta)$ error at ${\cal I}^+$.

\subsection*{Evolution equation for $J$}

In discretizing the evolution equation, we follow an approach that has
proven successful in the axisymmetric case~\cite{papa} and recast it in
terms of the 2-dimensional wave operator
\begin{equation}
\Box\,^{(2)}\psi=e^{-2 \beta}[2\psi_{,ru}-(\frac{V}{r} \psi_{,r})_{,r}]
\label{eq:2box}
\end{equation}
corresponding to the line element
\begin{equation}
d\sigma^2=2l_{(\mu}n_{\nu )}dx^\mu dx^\nu
=e^{2 \beta} du[\frac{V}{r} du +2 dr],
\label{eq:2metric}
\end{equation}
where $l_\mu =u_{,\mu}$ is the normal to the outgoing null cones and
$n_\mu$ is a null vector normal inwards to the spheres of constant
$r$.  Because the domain of dependence of $d\sigma^2$ contains the
domain of dependence induced in the $(u,r)$ submanifold by the full
space-time metric (\ref{eq:bmet}), this approach does not lead to
convergence problems.

The quasispherical evolution equation~(\ref{eq:wev}) then reduces to
\begin{equation}
e^{2\beta}\Box^{(2)}(rJ) = {\cal H},  \label{eq:wev1}
\end{equation}
where
\begin{equation}
{\cal H} = - r^{-1}(r^2 \eth U)_{,r}
+ 2 r^{-1} e^{\beta} \eth^2 e^{\beta} -(r^{-1}W)_{,r}J.
\end{equation}
Because all 2-dimensional wave operators are conformally flat, with
conformal weight $-2$, we can apply to (\ref{eq:wev1}) a flat-space
identity relating the values of $rJ$ at the corners $P$, $Q$, $R$ and
$S$ of a null parallelogram ${\cal A}$ with sides formed by incoming
and outgoing radial characteristics. In terms of $rJ$, this
relation leads to an integral form of the evolution equation,
\begin{equation}
\left(r J\right)_Q = \left(rJ\right)_P + \left(rJ\right)_S
- \left(rJ\right)_R + \frac{1}{2} \int_{\cal A} du\,dr {\cal H}.
\end{equation}

The corners of the null parallelogram cannot be chosen to lie exactly
on the grid because the velocity of light in terms of the $x$
coordinate is not constant. Numerical analysis and experimentation has
shown~\cite{Gom} that a stable algorithm results by placing this
parallelogram so that the sides formed by incoming rays intersect
adjacent $u$-hypersurfaces at equal but opposite $x$-displacement
from the neighboring grid points. The elementary computational cell
consists of the lattice points $(n,i,k,l)$ and $(n,i\pm 1,k,l)$ on the
"old" hypersurface and the points $(n+1,i,k,l)$, $(n+1,i-1,k,l)$ and
$(n+1,i-2,k,l)$.

The values of $rJ$ at the vertices of the parallelogram are
approximated to second order accuracy by linear interpolations between
nearest neighbor grid points on the same outgoing characteristic. Then,
by approximating the integrand by its value at the center $C$ of the
parallelogram, we have
\begin{equation}
\left(rJ\right)_Q = \left(rJ\right)_P + \left(rJ\right)_S - \left(rJ\right)_R
 + \frac{1}{2}\Delta u \left(r_Q - r_P + r_S - r_R\right) {\cal H}_C .
\label{eq:wev2}
\end{equation}
As a result, the discretized version of (\ref{eq:wev1}) is given by
\begin{equation}
\left(rJ\right)^{n+1}_i = {\cal{F}}
                          \left( \left(rJ\right)^{n+1}_{i-1},
                                 \left(rJ\right)^{n+1}_{i-2},
                                 \left(rJ\right)^{n}_{i+1},
                                 \left(rJ\right)^{n}_i,
                                 \left(rJ\right)^{n}_{i-1}
                          \right)
                        + \frac{1}{2} \Delta u
                          \left( r_Q - r_P + r_S - r_R \right) {\cal H}_C
\label{eq:wev3}
\end{equation}
where ${\cal{F}}$ is a linear function of the $(rJ)$'s and angular
indexes have been suppressed. Consequently, it is possible to move
through the interior of the grid computing $(rJ)^{n+1}_i$ by an
explicit radial march using the fact that the value of $rJ$ on the
world tube is known.

The above scheme is sufficient for second order accurate evolution in
the interior of the radial domain. However, for startup purposes,
special care must be taken to handle the second radial point.  In
determining $(rJ)^{n+1}_{i=2}$ the strategy (\ref{eq:wev2}) is easily
modified so that just two radial points are needed on the $n+1$ level;
the parallelogram is placed so that $P$ and $Q$ lie precisely on
$(n+1,1,i,j)$ and $(n+1,2,i,j)$ respectively. Note that the calculation
of ${\cal H}_C$ poses no problems, since the values of $W$, $U$, and
$U_{,r}$ are known on the worldtube and the value of $W_{,r}$ on the
worldtube can be calculated by (\ref{eq:ww}).

In order to apply this scheme globally we must also take into account
technical problems concerning the order of accuracy for points near
${\cal I}^+$.  For this purpose, it is convenient to renormalize
(\ref{eq:wev3}) by introducing the intermediate variable $\Phi = (rJ)
(1-x) = R x J$.  This new variable has the desired feature of finite
behavior at ${\cal I}^+$. With this substitution the evolution equation
becomes
\begin{equation}
\Phi_Q =
  \frac{1}{4} x_Q \Delta u {\cal H}_C
+ \frac{1 - x_Q} {1-x_P}
  \left(\Phi_P - \frac{1}{4} x_P \Delta u {\cal H}_C\right)
+ \frac{1 - x_Q} {1-x_S}
  \left(\Phi_S + \frac{1}{4} x_S \Delta u {\cal H}_C\right)
- \frac{1 - x_Q} {1-x_R}
  \left(\Phi_R + \frac{1}{4} x_R \Delta u {\cal H}_C\right)
\end{equation}
where all the terms have finite asymptotic value.

\section{Tests} \label{sec:4}

Some of the fundamental issues underlying stability of the evolution
algorithm are discussed in Appendix~\ref{app:stab}. We have carried out
numerical experiments which confirm that the code is stable, subject to
the CFL condition, in the perturbation regime where caustics and
horizons do not form. The first set of tests consist of evolving short
wavelength initial null data, with all world tube data set to zero. In
this case, the world tube effectively acts as a mirror to ingoing
gravitational waves. The tests were run until all waves were reflected
and radiated away to ${\cal I}^+$. In particular, data with $\mid J
\mid \approx 10^{-6}$ was run from from $u=0$ to $u=40$, corresponding
to approximately $10^4$ timesteps, at which time it was checked that
the amplitude was decaying.

In the second set of tests, we included short wavelength data with
amplitude $10^{-4}$ for the boundary values of $\beta$, $J$, $U$, $Q$
and $W$ on the world tube (with compact support in time) as well for
the initial data for $J$ (with compact support on the initial null
hypersurface).  Again the code was run for approximately $4500$
timesteps (from $u=0$ to $u=25$), at which time all fields were
decaying exponentially. This test reveals a favorably robust stability
of the worldtube initial value problem, since in this case the world
tube conservation conditions which guarantee that the exterior
evolution be a vacuum Einstein solution were not imposed upon the
worldtube data.

We now present code tests for the accuracy of numerical solutions and
their waveforms at infinity. The tests are based upon linearized
solutions on a Minkowski background and linearized Robinson-Trautman
solutions. These solutions provide testbeds for code calibration as
well as consistent worldtube boundary values for an external vacuum
solution. In addition, we use numerical solutions of the nonlinear
Robinson-Trautman equation to study the waveform errors introduced by
the quasispherical approximation.

\subsection{Linearized solutions}

Appendices \ref{app:gauge} and \ref{app:lin} describe how to generate
3-dimensional linearized solutions on a Minkowski background in null
cone coordinates and their gauge freedom. To calibrate the accuracy of
the code, we choose a solution of (\ref{eq:wave}) and (\ref{eq:Ecr})
which represents an outgoing wave with angular momentum $l=6$ of the
form
\begin{equation}
 \Phi =(\partial_z)^6\frac{1}{u^2r}, \label{eq:exacsol}
\end{equation}
where $\partial_z$ is the $z$-translation operator. The resulting
solution is well behaved above the singular light cone $u=0$.

Convergence was checked in the linearized regime by choosing initial
data of very small amplitude $(\mid J \mid \approx 10^{-9})$. We used
the linearized solution (\ref{eq:exacsol}) to give data at $u=1$, with
the inner boundary at $R=1$, and we compared the numerically evolved
solution at $u=1.5$. The computation was performed on grids of size
$N_x$ equal $128$, $192$, $256$ and $320$, while keeping $N_x=4
N_\zeta$. Convergence to second order was verified in
the $L_1$, $L_2$ and $L_{\infty}$ norms.

\subsection{Robinson-Trautman solutions}

The Robinson-Trautman space-times~\cite{RT} contain a distorted black
hole emitting purely outgoing radiation. The metric can be put in the
Bondi form
\begin{equation}
  ds^2 = -({\cal K}-{2\over r{\cal W}})du^2-2{\cal W}dudr
         -2r{\cal W}_{,A}dudx^A+r^2q_{AB}dx^A dx^B,
\end{equation}
where $K={\cal W}^2[1-L^2(\log {\cal
W})]$, $L^2$ is the angular momentum operator and ${\cal W}(u,x^A)$
satisfies the nonlinear equation
\begin{equation}
      12 \partial_u(\log {\cal W}) = {\cal W}^2 L^2 {\cal K}.
    \label{eq:rteq}
\end{equation}

The Schwarzschild solution (for a unit mass black hole) is obtained
when ${\cal W}=1$. More generally, smooth initial data ${\cal
W}(u_0,x^A)$ evolves smoothly to form a Schwarzschild horizon. The
linearized solutions to the Robinson-Trautman equation (\ref{eq:rteq})
are obtained by setting ${\cal W}=1+\phi$ and dropping nonlinear terms
in $\phi$:
\begin{equation}
    12 \partial_u \phi = L^2(2-L^2)\phi.  \label{eq:rt}
\end{equation}
For a spherical harmonic perturbation $\phi =A(u)Y_{\ell m}$ this leads
to the exponential decay $A=A(0)e^{- u
\,\ell(\ell+1)(\ell^2+\ell-2)/12}$.

These linearized solutions provide analytic worldtube data for our
evolution code, along with the initial null data $J=0$. We have used
this as a check of code accuracy in the perturbative regime off
Schwarzschild. With this data, the code should evolve $J$ to be
globally zero to second order in grid size.  Of particular importance
for the extraction of waveforms, this should hold for the value of $J$
at ${\cal I}^+$. We have carried out such a test with a small
extraction radius ($R=3m$) and a linearized solution of the form:
\begin{equation}
{\cal W} = 1 + \lambda \, \Re[( e^{-2u} Y_{22} + e^{-10u} Y_{33} )]
\end{equation}
with $\lambda=10^{-5}$.
The error norm
\begin{equation}
||{\cal E}_J||^2=\int_0^{u_1} du \int d\Omega J^2
\end{equation}
was determined by integration over a sphere at $\cal I^+$ with solid
angle element $d\Omega$, and with an integration time of $u_1=2$. The
convergence rate to the true value was found to be $O(\Delta^{1.92})$.

We have also obtained second-order accurate numerical solutions to the
nonlinear Robinson-Trautman equation (\ref{eq:rt}). See Ref.~\cite{eth}
for numerical details. This allows us to check the discrepancy between
exact waveforms and waveforms obtained by regarding the whole spacetime
in the quasispherical perturbative approximation. We have based this
comparison on initial data in modes
\begin{equation}
  {\cal W}|_{u=0}=1+\lambda{\Re}[Y_{lm}].
\end{equation}
In order to supply some physical perspective, the nonlinearity of the
initial data is best measured in terms of $\epsilon = \left((M-M_S)/M
\right)^{1/2}$, where $M$ is the initial mass of the system and $M_S$
is the mass of the corresponding Schwarzschild background. (Here,
$M_S=1$). We also calculate the percentage of the initial mass which is
radiated away during the entire course of our simulations. The Bondi
news function determines the mass loss and it is also an appropriate
physical quantity to invariantly describe radiative waveforms. In the
coordinates adopted here, the news function is given by~\cite{robjef}
\begin{equation} N(u_B,x^A)=  \frac{1}{2}\ {\cal W}^{-1}\eth^2 {\cal W}
\end{equation} where the Bondi time $u_B$ measured by observers at
${\cal I^+}$ is related to $u$ by $du_B/du ={\cal W}$.

\begin{figure}
\centerline{\epsfysize=576pt\epsfbox{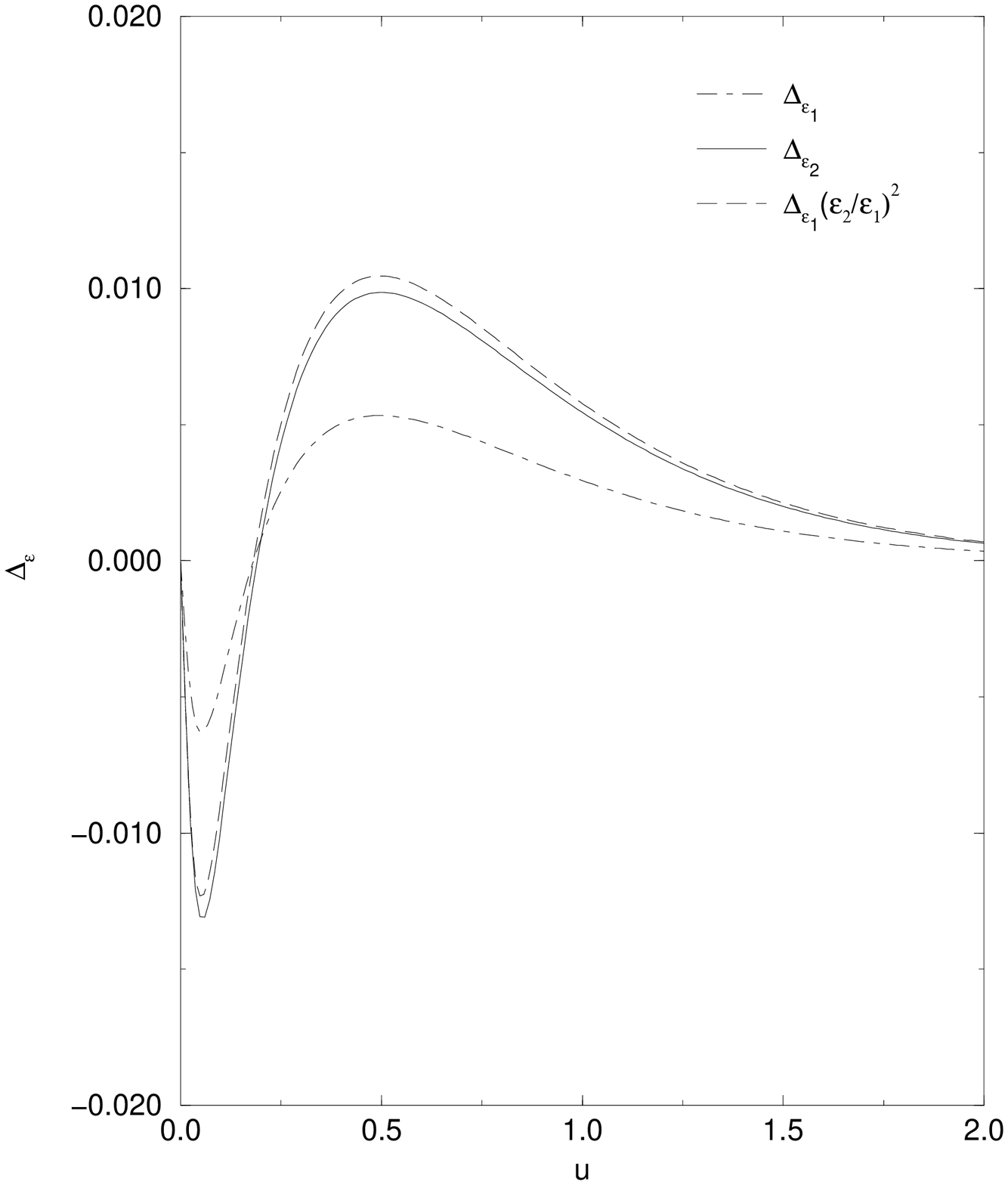}}
\caption{$\Delta_{\epsilon}$ for $\epsilon_1=0.14$ and $\epsilon_2=0.22$
(corresponding to a total mass loss of $0.6\%$ and $1.2\%$ respectively) for
initial data in a $Y_{22}$ mode. In this regime $\Delta_{\epsilon}$
scales as $\epsilon^2$, thus indicating that first order perturbation is
valid in this regime. }
\label{fig:kuabi_2}
\end{figure}

\begin{figure}
\centerline{\epsfysize=576pt\epsfbox{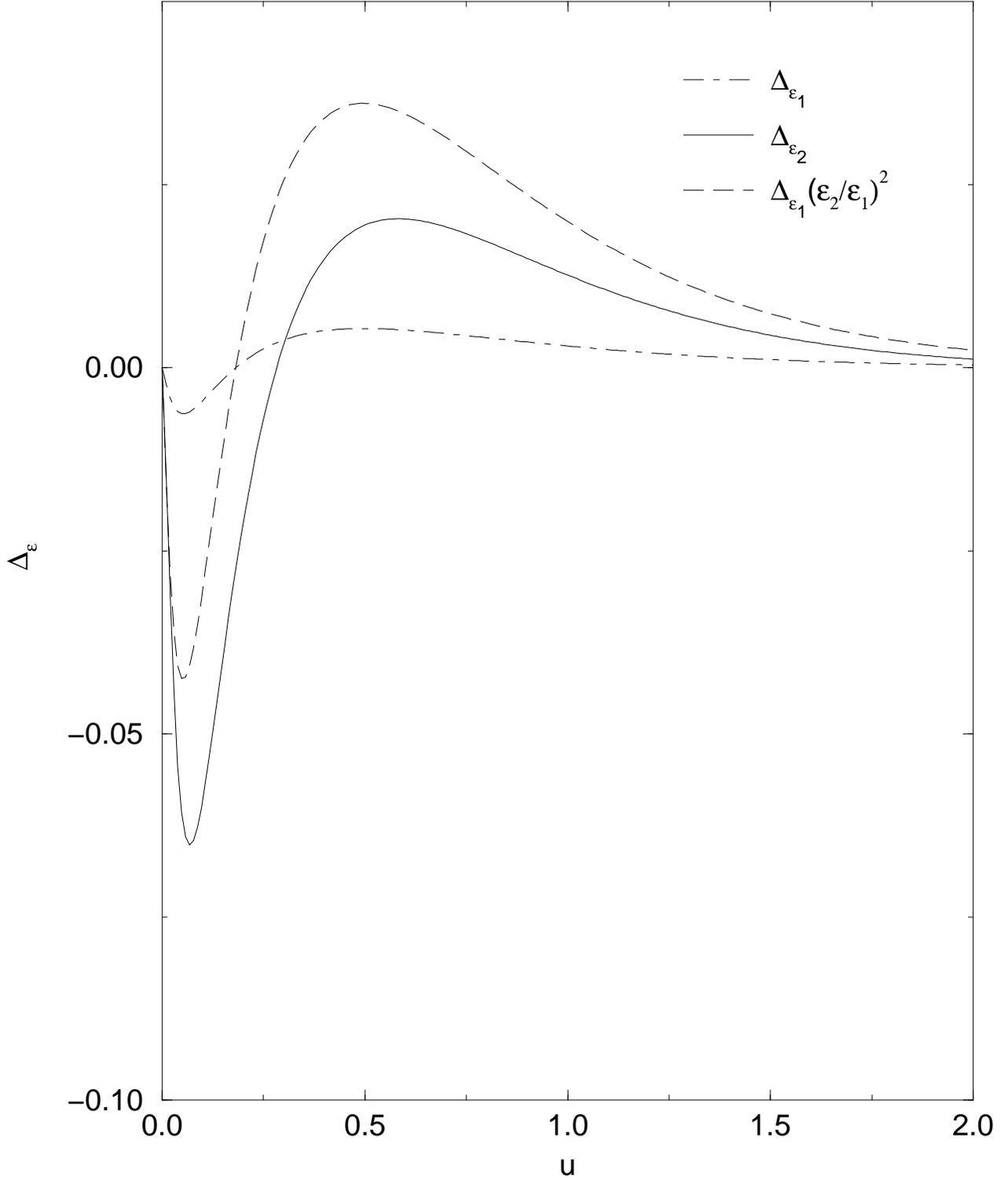}}
\caption{$\Delta_{\epsilon}$ for $\epsilon_1=0.14$ and $\epsilon_2=0.4$
(corresponding to a total mass loss of $0.6\%$ and $4.6\%$
respectively) for initial data in a $Y_{22}$ mode.  The
difference between quadratically rescaling $\Delta_{\epsilon}$ and its
actual value is about $40\%$, indicating that second order perturbation
is inaccurate in this regime. }
\label{fig:kuama_2}
\end{figure}

For various initial modes, we have calculated the news function for the
numerical solution of the nonlinear R-T equation ( $N^n_{\epsilon}$ )
and compared it with the news function of the linearized solution
($N^p_{\epsilon}$). As expected, for small values of $\epsilon$ they
agree up to second order in $\epsilon$. Figure \ref{fig:kuabi_2} graphs
the time dependence of $\Delta_{\epsilon} = N^n_{\epsilon} -
N^p_{\epsilon}$ (for a representative angle) for a system initially in
a $l=2$, $m=2$ mode, which is the dominant gravitational radiation mode
for a spiraling binary system. The figure illustrates that
$\Delta_{\epsilon}$ scales with $\epsilon^2$.  However, for larger
$\epsilon$, corresponding to a total radiative mass loss greater than
$2.5\%$, this is no longer the case and a noticeable discrepancy
arises. For instance, as illustrated in Figure \ref{fig:kuama_2}, the
difference between quadratically rescaling $\Delta_{\epsilon}$ and its
actual value is about $40\%$ for a mass loss of $4\%$.

\begin{figure}
\centerline{\epsfysize=576pt\epsfbox{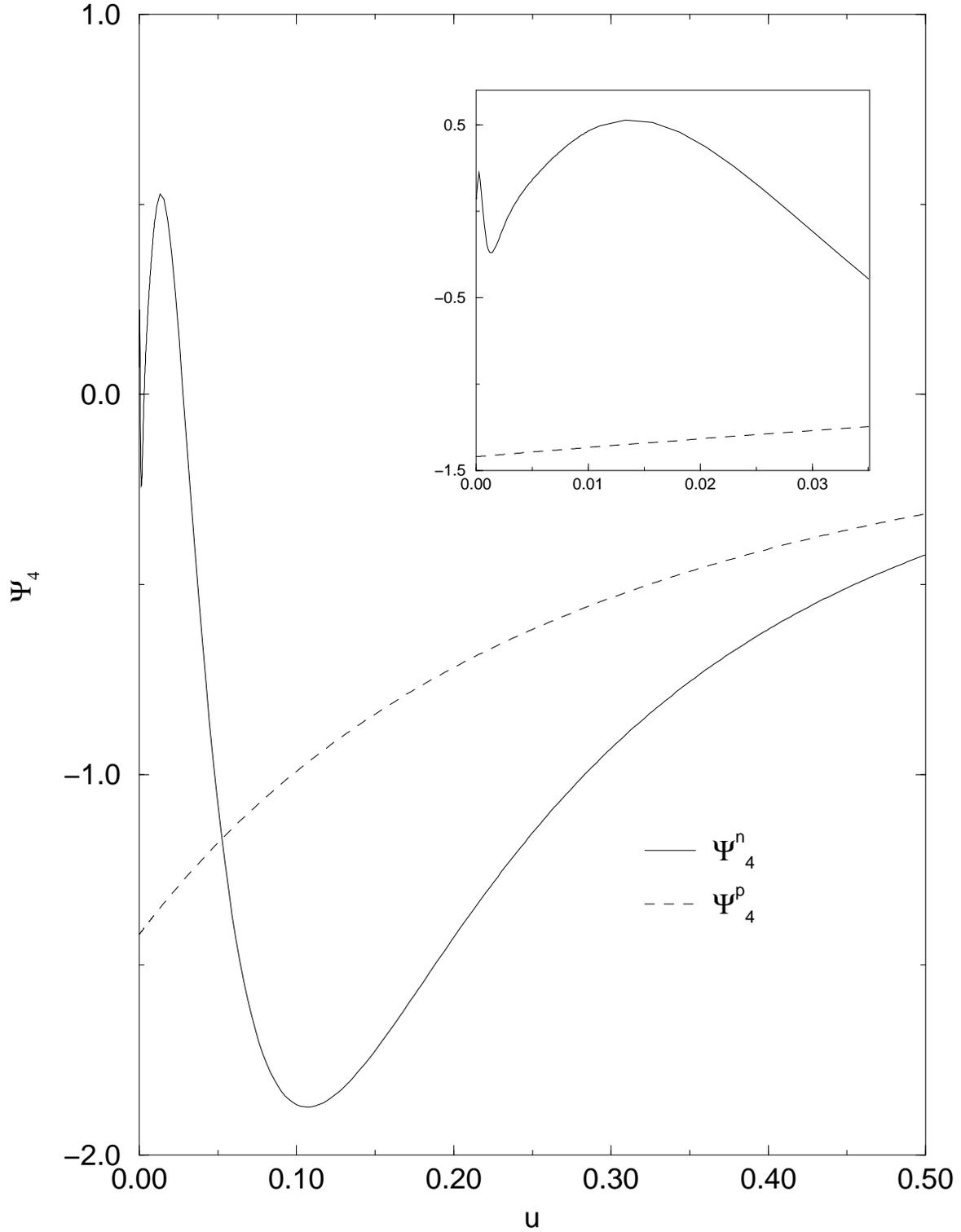}}
\caption{ $\Psi^n_4$ and $\Psi^p_4$ for a point lying $10$ degrees
above the equator and initial data in a $Y_{22}$ mode. The total
mass loss is $4\%$.  The insert shows the marked oscillatory behavior
at early times. }
\label{fig:radia}
\end{figure}

Hence, this indicates that not only the first order perturbation
treatment but also the second order treatment is grossly inaccurate in
this regime.  Serious discrepancies arise between $N^n_{\epsilon}$ and
$N^p_{\epsilon}$ for ranges in which the mass loss is not extreme. In
fact, $N^n_{\epsilon}$ reveals an oscillatory behavior qualitatively
quite different from the pure decaying mode of the perturbative
solution, which has serious implications for the tidal acceleration
which the radiation would produce in a distant gravitational wave
antenna. As measured by the radiative component of the Weyl tensor
$\Psi_4$, the tidal acceleration is given by the Bondi-time derivative
of the news function. In contrast to the monotonic decay of the
perturbative solution, the actual behavior of $\Psi_4$ exhibits damped
oscillations. For a $Y_{22}$ initial mode, Figure \ref{fig:radia} shows
the drastic difference between the numerically obtained $\Psi^n_4$ and
the corresponding $\Psi^p_4$ calculated with the perturbative solution.

Similar nonlinear oscillations arise from other choices of initial
data. Some partial explanation of this behavior might be possible using
second order perturbation theory for the Robinson-Trautman equation
{}~\cite{fritmor} but the full behavior would require perturbation
expansions far beyond practicality.

\section{Computational efficiency of CCM} \label{sec:5}

This section is concerned with the computational efficiency of a
numerical calculation of gravitational radiation from an isolated
system, such as binary black hole. By ``computational efficiency'' we
mean the amount of computation $A$ (i.e. the number of floating point
operations) for a given target error $\varepsilon$.  We will show
that the computational efficiency of the CCM algorithm is never
significantly worse than that of a pure Cauchy algorithm; and that for
high accuracy the CCM algorithm is always much more efficient.

In CCM a ``3 + 1'' interior Cauchy evolution is matched to an exterior
characteristic evolution at a worldtube $r_M = constant$. A key feature
is that the characteristic evolution can be rigorously compactified, so
that the whole space-time to future null infinity may be represented on
a finite grid. From a numerical point of view this means that the only
error made in a calculation of the gravitational radiation at infinity
is due to the finite discretization $\Delta$;  for second-order
algorithms this is $O(\Delta^2)$.  The value of the matching radius
$r_M$ is important and it will turn out that, for efficiency, it should
be as small as possible. However, caustics may form if $r_M$ is too
small. The smallest value of $r_M$ that avoids caustics is determined
by the physics of the problem, and is {\em not} affected by either the
discretization $\Delta$ or the numerical method.

On the other hand, the standard approach is to make an estimate of the
gravitational radiation solely from the data calculated in a pure
Cauchy evolution. The simplest method would be to use the raw data, but
that approach is too crude because it mixes gauge effects with the
physics. Thus a substantial amount of work has gone into perturbative
methods that factor out the gauge effects using multipole expansions
and estimate the gravitational field at infinity from its behavior
within the domain of the Cauchy computation~\cite{ab1,ab2,ab3}. We will
call this method waveform extraction, or WE.  While WE is a substantial
improvement on the crude approach, it ignores the nonlinear terms in
the Einstein equations. The resulting error will be estimated below.

Both CCE and WE are {\em extraction} methods. That is, they use Cauchy
data on a worldtube $\Gamma$ to estimate gravitational waveforms at
infinity, and they have no  back effect on the Cauchy evolution.  In
both methods there is an error (which is difficult to estimate) due to
the artificial Cauchy outer boundary condition. The difference between
CCE and WE is in the treatment of the nonlinear terms between $\Gamma$
and future null infinity and in the truncation of the perturbative
multipole expansion at some low order.  WE ignores the nonlinear terms,
and this is an inherent limitation of a perturbative method. Even if it
is possible to extend WE beyond linear order, there would necessarily
be a cut-off at some finite order. The quasispherical implementation of
CCE incorporates all multipole contributions but also ignores the
nonlinear terms. However, it is in principle straightforward to
incorporate these terms into the code. A full implementation of CCE
would do so, and the nonlinear terms would be treated without error.

\subsection{Error estimate in WE}

We assume that a pure Cauchy evolution proceeds in a spatial
domain of radius $r_B$, and the extraction is computed on a worldtube
$\Gamma$ of radius $R$, with $R<r_B$.

The evolution equation (\ref{eq:wev}) may be written:
\begin{equation}
(rJ)_{,ur} = quasispherical \ part + \frac{1}{2} N_J
\label{eq:WE1}
\end{equation}
with the nonlinear term $N_J$ given by (\ref{eq:nj}) (Actually, $N_J$
also implicitly contains contributions from $(\int N_Q dr)/r^2$ and
$\int N_U dr$, and from the quasispherical approximation of terms in
(\ref{eq:wev}), but these effects are all of the same order as $N_J$).
The order of magnitude of various terms can be expressed in terms of a
function $c(u,x^A)$ (whose time-derivative is the news function); note
that $c$ is not a small quantity. The expressions are:
\begin{eqnarray}
J=O(\frac{c}{r}),\, h_{AB}-q_{AB}=O(\frac{c}{r}),\,
h_{AB_{,r}}=O(\frac{c}{r^2}),\,
\beta=O(\frac{c^2}{r^2}),\nonumber \\
Q=O(\frac{c}{r}),\, U=O(\frac{c}{r^2}),\,
C^C_{AB}=O(\frac{c}{r}),\, W=O(\frac{c^2}{r^2}).
\label{eq:WE2}
\end{eqnarray}
These estimates are obtained by the radial integration of the field
equations in Sec. \ref{sec:2.2}, assuming that the background
geometry is Minkowskian and that the Bondi gauge conditions are
satisfied. Should this not be the case then constants of order unity
would be added to $Q$, $U$ and $W$, and the effect of this would be to
amend (\ref{eq:wev}) by adding terms to the quasispherical part so that
it represents wave propagation on a (fixed) non-flat background.
However, the order of magnitude of terms in the nonlinear part would
not be affected.  Thus there is no loss of generality, and a
significant gain in simplicity and transparency, in performing the
error analysis on a Minkowskian background.

It is straightforward to confirm that the nonlinear correction to
(\ref{eq:wev}) involves terms of order $O(c^2/r^3)$ or smaller.
WE estimates the waveform at future null infinity from data at $r=R$.
This could be made exact (modulo the error introduced by truncating the
multipole expansion) if the nonlinear part of (\ref{eq:wev}) were zero.
Thus the error introduced by ignoring $N_J$ is
\begin{equation}
\varepsilon(c_{,u}) \equiv (c_{,u})_{exact}-(c_{,u})_{WE} =
\int_{R}^{\infty}O(\frac{c^2}{r^3}) dr = O(\frac{c^2}{R^2}).
\label{eq:WE4}
\end{equation}
In the case of the collision of two black holes, with total mass $M$
and with $c=O(M)$, the error is $O(M^2/R^2)$ and it is tempting to say
that if extraction is performed at $R=10 M$ then the expected error of
the WE method is about 1\%. This would be quite wrong because there is
no reason for the constant factor in $O(M^2/R^2)$ to be approximately~$1$.

\subsection{Computational efficiency}

A numerical calculation of the emission of gravitational radiation
using a CCM algorithm is expected to be second-order convergent, so
that after a fixed time interval the error
\begin{equation}
\varepsilon = O(\Delta^2) \simeq k_1 \Delta^2,
\label{eq:CE1}
\end{equation}
where $\Delta$ is the discretization length and $k_1$ is a constant.
On the other hand, the same calculation using WE must allow for the
error found in (\ref{eq:WE4}), and therefore after the same fixed time
interval there will be an error of:
\begin{equation}
\varepsilon=O(\Delta^2,R^{-2}) \simeq max(k_2\Delta^2,\frac{k_3}{R^2}),
\label{eq:CE2}
\end{equation}
where $k_2$ and $k_3$ are constants.

We now estimate the amount of computation required for a given desired
accuracy.  We make one important assumption:
\begin{itemize}
\item The computation involved in matching, and in waveform extraction,
is an order of magnitude smaller than the computation involved in
evolution, and is ignored. This is justified by the 2D nature of the
extraction and matching processes as opposed to the 3D nature of
evolution.
\end{itemize}
For the sake of transparency we make some additional simplifying
assumptions (otherwise some extra constants of order unity would appear
in the formulas below but the qualitative conclusions would be
unaffected):
\begin{enumerate}
\item The amount of computation per grid point per time-step, $a$, is the same
for the Cauchy and characteristic algorithms.
\item  The constants $k_1, k_2$ in the equations above are approximately
equal and will be written as $k$.
\item In CCM, the numbers of Cauchy and characteristic grid-points
are the same; thus the total number of grid points per time-step is
\begin{equation}
     \frac{8 \pi r_M^3}{3 \Delta^3}.
\label{eq:CE3}
\end{equation}
\item In WE, the outer boundary $r_B$ is at $\sqrt[3]{2} \, R$;
thus the total number of grid points per time-step is
\begin{equation}
   \frac{8 \pi R^3}{3 \Delta^3}.
\label{eq:CE4}
\end{equation}
\end{enumerate}
It follows that the total amount of computation $A$ required for the
two methods is:
\begin{equation}
A_{CCM}=\frac{8 \pi r_M^3 a}{3 \Delta^4},\ \ A_{WE}=
\frac{8 \pi R^3 a}{3 \Delta^4}.
\label{eq:CE5}
\end{equation}
Thus the method which requires the least amount of computation is
determined by whether $r_M > R$ or $r_M < R$. (Because of the
assumptions (1) to (4) this criterion is not exact but only
approximate.)

As stated earlier, the value of $r_M$ is determined by the physics,
specifically by the condition that the nonlinearities outside $r_M$
must be sufficiently weak so as not to induce caustics. The value of
$R$ is determined by the accuracy condition (\ref{eq:CE2}), and also by
the condition that the nonlinearities outside $R$ must be sufficiently
weak for the existence of a perturbative expansion. Thus we never
expect $R$ to be significantly smaller than $r_M$, and therefore the
computational efficiency of a CCM algorithm is never expected to be
significantly worse than that of a WE algorithm.

If high accuracy is required, the need for computational efficiency
always favors CCM.  More precisely, for a given desired error
$\varepsilon$, Eq's. (\ref{eq:CE1}) and (\ref{eq:CE2}) and assumption
(2) imply
\begin{equation}
\Delta=\sqrt{\varepsilon/k},\ R=\sqrt{k_3/\varepsilon}.
\label{eq:CE6}
\end{equation}
Thus
\begin{equation}
A_{CCM}=\frac{8 \pi r_M^3 a k^2}{3 \varepsilon^2},
\ A_{WE}=\frac{8 \pi a k^2 k_3^{3/2}}{3 \varepsilon^{7/2}},
\label{eq:CE7}
\end{equation}
so that
\begin{equation}
\frac{A_{CCM}}{A_{WE}}=\frac{r_M^3 \varepsilon^{3/2}}{k_3^{3/2}}
\rightarrow 0 \ as \ \varepsilon \rightarrow 0.
\label{eq:CE8}
\end{equation}
This is the crucial result: the computational intensity of CCM relative
to that of WE goes to zero as the desired error $\varepsilon$ goes to
zero.

\section{Conclusion}

The computer code described in this paper is a partial implementation
of CCE. That is, given data on an $r=constant$ worldtube $\Gamma$, the
code calculates the gravitational radiation at future null infinity in
the quasispherical approximation. A full implementation of CCE is
currently being developed which addresses the following issues:
\begin{itemize}
\item The ignored nonlinear terms in the Einstein equations must be
calculated, discretized and incorporated into the code.
\item Algorithms need to be developed to translate numerical Cauchy data
near $\Gamma$ into characteristic data on $\Gamma$.
\item In general $\Gamma$ will be described in terms of Cauchy
coordinates, and will not be exactly $r = constant$; the characteristic
algorithm needs amendment to allow for this.
\end{itemize}

Once a fully nonlinear CCE code has been achieved it will be possible,
under certain circumstances, to obtain second-order convergence to the
true analytic solution at future null infinity. For example, if
$\Gamma$ has radius $R$ and the radius of the Cauchy domain is $r_B$
($>R$), then causality implies that the gravitational field at $\Gamma$
will not be contaminated by boundary errors until time $t_C \approx
(r_B - R)$, where $t=0$ at the start of the simulation.  There is no
analytic error in the characteristic computation, so there will be no
analytic error in the gravitational radiation at future null infinity
for the initial time period $t_C$; under some circumstances this may be
the time period that is physically interesting.  Further, this time
period $t_C$ may be extended by using results from the characteristic
computation to provide the outer boundary condition in the Cauchy
calculation. This would amount to a partial implementation of CCM since
there would be data flow in both Cauchy to characteristic, and
characteristic to Cauchy, directions (The implementation is only
partial because $R$ and $r_B$ are very different). Since the data flow
is two-way, the possibility of a numerical instability arises. However,
the timescale of the growth of any instability would be $t_C$, and
therefore such a computation could be safely run for a time of several
$t_C$; the results obtained would be second-order
convergent to the true analytic solution.

Once the technology for Cauchy to characteristic, and characteristic to
Cauchy, data flow across an arbitrary worldtube has been developed, a
full implementation of CCM will amount to taking the limit in which the
outer boundary approaches the extraction worldtube. We are encouraged
to believe that this is feasible, i.e. without numerical instability,
because CCM {\em has} been achieved for the model problem of the
nonlinear 3D scalar wave equation~\cite{Ccpaper}.

\acknowledgements

This work was supported by the Binary Black Hole Grand Challenge
Alliance, NSF PHY/ASC 9318152 (ARPA supplemented), and by NSF PHY
9510895 to the University of Pittsburgh. N.T.B. thanks the South
African FRD for financial support, and the University of Pittsburgh for
hospitality during a sabbatical. Computer time has been provided by the
Pittsburgh Supercomputing Center under grant PHY860023P and by the High
Performance Computing Facility of the University of Texas at Austin.

\appendix
\section{Gauge Freedom}\label{app:gauge}

Given a metric in a Bondi null coordinate system, the gauge freedom is
\begin{equation}
\delta g^{ab} = g^{ac}\partial_c \xi^b +g^{cb}\partial_c \xi^a
-\xi^c \partial_c g^{ab}
\end{equation}
subject to the conditions $\delta g^{00}=0$, $\delta g^{0A}=0$ and
$g_{AB}\delta g^{AB}=0$. These latter conditions imply the functional
dependencies
\begin{equation}
\xi^0 = \kappa(u,x^B)
\end{equation}
\begin{equation}
\xi^A=f^A(u,x^B) - \int dr g_{01}g^{AB}\partial_B \kappa
\end{equation}
and
\begin{equation}
\xi^1= (r/2)(g_{01}g^{1A}\partial_A \kappa-D_B \xi^B).
\end{equation}
For a spherically symmetric background metric we drop quadratically
asymmetric terms to obtain
\begin{equation}
\xi=\eth \phi - e^{2\beta}r^{-1}\eth \kappa
\end{equation}
and
\begin{equation}
\xi^1 = - {r\over 4} \left( \bar \eth \xi + \eth \bar \xi \right)
      =   {1\over 4} \left[ - r \eth \bar \eth \left( \phi + \bar \phi \right)
                            + 2 e^{2\beta} \eth \bar \eth \kappa \right],
\end{equation}
where $q_A\xi^A=\xi$ and $q_Af^A=\eth \phi$, in terms of a complex
scalar field $\phi(u,x^A)$.

This gives rise to the following gauge freedom in the metric quantities:
\begin{equation}
    \delta J = -\eth^2 \phi + { {e^{2\beta}} \over r} \eth^2 \kappa,
\end{equation}
\begin{equation}
 \delta e^{2\beta} = - \left(e^{2\beta} \kappa \right)_{,u}
                     - e^{2\beta} \xi^1_{,r}
                   = - \left(e^{2\beta} \kappa\right)_{,u}
                     + \frac{1}{4} e^{2\beta} \eth \bar \eth(\phi+\bar \phi),
\end{equation}
\begin{equation}
   \delta U = \xi_{,u} - \frac{V}{r} \xi_{,r}
            - { {e^{2\beta}} \over {r^2} } \eth \xi^1
\end{equation}
and
\begin{equation}
 \delta V = - (2r\xi^1 + \kappa V)_{,u} + V \xi^1_{,r}
            - r \xi^1 \left( \frac{V}{r} \right)_{,r}.
\end{equation}

\section{Linear Solutions}\label{app:lin}

We present a 3D generalization of a scheme~\cite{papa} for generating
linearized solutions off a Minkowski background in terms of spin-weight
0 quantities $\alpha$ and $Z$, related to $J$ and $U$ by $J=2\eth^2
\alpha$ and $U=\eth Z$. We may in this approximation choose a gauge in
which $\beta=0$ or otherwise use the gauge freedom to set
$\beta=H(u,x^A)$. In either case, $W$ is given by the radial
integration of the linearization of (\ref{eq:ww}) and the remaining
linearized equations reduce to
\begin{equation}
 (r^4  Z_{,r})_{,r}  =  -2 r^2(2-L^2)\alpha_{,r}
  \label{eq;hz}
\end{equation}
 and
\begin{equation}
E := 2 \left( r \alpha \right)_{,ur}
   - \frac{1}{r} \left( r^2 \alpha_{,r} - \frac{1}{2} r^2 Z \right)_{,r} = 0,
\label{eq:E}
\end{equation}
where $L^2= -\eth \bar \eth$ is the angular momentum operator.

Now set
\begin{equation}
   r^2 \alpha_{,r}= \left( r^2 \Phi \right)_{,r}
   \label{eq:alpha}
\end{equation}
and
\begin{equation}
   r^2 Z_{,r}= 2 \left(L^2 -2 \right) \Phi.
   \label{eq:Z}
\end{equation}
Then
\begin{equation}
  E = r\Box \, \Phi + 2 \left(\Phi +\alpha\right)_{,u}
    - \frac{2}{r^{2}} \left(r^2 \Phi\right)_{,r} + Z,
\end{equation}
where $\Box$ is the wave operator
\begin{equation}
   r \Box\, \Phi = 2 \left(r \Phi \right)_{,ur}
   - \left(r \Phi \right)_{,rr} + \frac{1}{r} L^2 \Phi .
   \label{eq:wave}
\end{equation}
It follows that
\begin{equation}
  E_{,r}= \frac{1}{r^2} \left(r^3 \Box \, \Phi \right)_{,r}.
  \label{eq:Ecr}
\end{equation}

Suppose now that $\Phi$ is a complex solution of the wave equation
$\Box \Phi =0$.  Then Eq. (\ref{eq;hz}) is satisfied as a result of
(\ref{eq:alpha}) and (\ref{eq:Z})and (\ref{eq:Ecr}) implies
$E_{,r}=0$.  If $\Phi$ is smooth and $O(r^2)$ at the origin, this
implies $E=0$, so that the linearized equations are satisfied
globally.  The condition that $\Phi=O(r^2)$ eliminates fields with only
monopole and dipole dependence so that it does not restrict the
generality of the spin-weight 2 function $J$ obtained.  Any global,
asymptotically flat linearized solution may be generated this way.

Alternatively, given a wave solution $\Phi$ with possible singularities
inside some worldtube, say $r=R$, we may generate an exterior solution,
corresponding to radiation produced by sources within the worldtube, by
requiring $E|_R=0$ or
\begin{equation}
 \left( \left(\Phi+\alpha \right)_{,u}
       - \frac{1}{r^2} \left(r^2 \Phi \right)_{,r}
       + \frac{1}{2} Z \right) |_R = 0.
\end{equation}
This is a constraint on the integration constants obtained in
integrating (\ref{eq:alpha}) and (\ref{eq:Z}) which may be satisfied by
taking $Z|_R=0$ and
\begin{equation}
\alpha_{,u}|_R = \left( \frac{1}{r^2} \left(r^2 \Phi \right)_{,r}
                        - \Phi_{,u} \right) |_{R}.
\end{equation}
This determines an exterior solution in a gauge such that $U|_R=0$.

\section{Linear Stability Analysis}\label{app:stab}

In the characteristic formulation, the linearized equations form the
principle part of the full system of Bondi equations. Therefore insight
into the stability properties of the full evolution algorithm may be
obtained at the linearized level. Here we sketch the von Neumann
stability analysis of the algorithm for the linearized Bondi equations,
generalizing a previous treatment given for the axisymmetric case. The
analysis is based up freezing the explicit functions of $r$ and
stereographic coordinate $\zeta$ that appear in the equations, so that
it is only valid locally for grid sizes satisfying $\Delta r << r$ and
$|\Delta \zeta|<<1$. However, as is usually the case, the results are
quite indicative of the stability of the actual global behavior of the
code.

Setting $G=rJ$ and $\Gamma=r^2U$ and freezing the explicit factors of
$r$ and $\zeta$ at $r=R$ and $\zeta=0$, the linearization of the Bondi
equations (\ref{eq:wq}), (\ref{eq:wu}) and (\ref{eq:wev}) takes the
form
\begin{equation}
   R^2 \Gamma_{,rr} - 2 \Gamma = - \left(R\, G_{,r} - G \right)_{,\zeta}
  \label{eq:gamma}
\end{equation}
and
\begin{equation}
  2\, G_{,ur} - G_{,rr} = - \frac{1}{R} \Gamma_{,r\bar \zeta}.
  \label{eq:bigg}
\end{equation}
Writing $\zeta=s_1+is_2$, introducing the Fourier modes
$G=e^{wu}e^{ikr}e^{il_1s_1}e^{il_2s_2}$ (with real $k$, $l_1$ and
$l_2$) and setting $\Gamma=AG$, these equations imply
\begin{equation}
  A = -i \left(1 - i k R \right) \left(l_1 - i l_2 \right)
    / \left[ 2 \left( 2 + k^2 R^2 \right) \right]
\end{equation}
and
\begin{equation}
  2 w = i k - \left(l_1^2 + l_2^2 \right)
              \left(1 - i k R \right)
              / \left[ (4R) \left(2 + k^2 R^2 \right) \right],
\end{equation}
representing damped quasinormal modes.

Consider now the FDE obtained by putting $G$ on the grid points $r_I$
and $\Gamma$ on the staggered points $r_{I+1/2}$, while using the same
stereographic grid $\zeta_J$ and time grid $u_N$. Let $P$, $Q$, $R$ and
$S$ be the corner points of the null parallelogram algorithm, placed so
that $P$ and $Q$ are at level $N+1$, $R$ and $S$ are at level $N$, and
so that the line $PR$ is centered about $r_I$ and $QS$ is centered
about $r_{I+1}$. Then, using linear interpolation and centered
derivatives and integrals, the null parallelogram algorithm for the
frozen version of the linearized equations leads to the FDE's

\begin{equation}
  \left( \displaystyle{{R \over {\Delta r}}} \right)^2
  \left(  \Gamma_{I+\frac{3}{2}} - 2 \Gamma_{I+\frac{1}{2}}
        + \Gamma_{I-\frac{1}{2}} \right)
- \left( \Gamma_{I+\frac{3}{2}} + \Gamma_{I-\frac{1}{2}} \right)
= - \delta_{\zeta} \left[ \displaystyle{{R \over {\Delta r}}}
       \left( G_{I+1} - G_{I} \right)
       - \frac{1}{2} \left(G_{I+1}+G_{I} \right) \right]
\label{eq:fgamma}
\end{equation}
(all at the same time level) and
\begin{eqnarray}
  G_{I+1}^{N+1}-G_I^{N+1}-G_{I+1}^N+G_I^N
  + { {\Delta u} \over {4 \Delta r}}
 \left( - G_{I+1}^{N+1} + 2\, G_I^{N+1} - G_{I-1}^{N+1}
        - G_{I+2}^N     + 2\, G_{I+1}^N - G_I^N \right)
  \nonumber \\
 = - { {\Delta u} \over {4 R}} \delta_{\bar \zeta}
      \left (  \Gamma_{I+\frac{1}{2}}^{N+1} - \Gamma_{I-\frac{1}{2}}^{N+1}
             + \Gamma_{I+\frac{3}{2}}^N     - \Gamma_{I+\frac{1}{2}}^N \right),
        \label{eq:fbigg}
\end{eqnarray}
where $\delta_{\zeta}$ represents a centered first derivative. Again
setting $\Gamma=AG$ and introducing the discretized Fourier modes
$G=e^{wN\Delta u}e^{ikI\Delta r}e^{il_1J_1\Delta s_1}e^{il_2J_2\Delta
s_2}$, we have $\delta_{\zeta}=L$ and $\delta_{\bar \zeta}=- \bar L$,
where $L=(1/2)[\sin(l_2\Delta s_2)/(\Delta s_2)+i\sin(l_1\Delta
s_1)/(\Delta s_1)]$, and (\ref{eq:fgamma}) and (\ref{eq:fbigg}) reduce
to
\begin{equation}
   A \left[ \left( {R \over {\Delta r}} \right)^2
            \left( 1 - \cos\alpha \right) + \cos\alpha \right] =
   L \left[ i { R \over {\Delta r}} \sin \left( \frac{\alpha}{2} \right)
            - \frac{1}{2} \cos \left( \frac{\alpha}{2} \right) \right]
  \label{eq:ffgamma}
\end{equation}
and
\begin{equation}
  e^{w\Delta u} = - e^{i\alpha} \left( { {\bar C - AD} \over {C-AD} } \right),
  \label{eq:stab}
\end{equation}
where $\alpha =k\Delta r$, $C=ie^{i\alpha/2}\sin(\alpha/2)+(\Delta
u/4\Delta r)(1-\cos\alpha)$ and $D=(i\bar L\Delta u/4R)\sin(\alpha/2)$.
The stability condition that $Re(w)\le 0$ then reduces to $Re[C(AD-\bar
A \bar D)]\ge 0$. It is easy to check that this is automatically
satisfied.

As a result, local stability analysis places no constraints on the
algorithm. It may seem surprising that no analogue of a
Courant-Friedrichs-Levy (CFL) condition arises in this analysis. This
can be understood in the following vein. The local structure of the
code is implicit, since it involves 3 points at the upper time level.
The stability of an implicit algorithm does not necessarily require a
CFL condition. However, the algorithm is globally explicit in the way
that evolution proceeds by an outward radial march from the origin. It
is this feature that necessitates a CFL condition in order to make the
numerical and physical domains of dependence consistent. In practice
the code is unstable unless the  domain of dependence determined by the
characteristics is contained in the numerical domain of dependence.  It
is important to note that if $U$ (or $\Gamma$) are not discretized on a
staggered grid then the above analysis shows the resulting algorithm to
be unconditionally unstable regardless of any CFL condition.


\end{document}